\documentclass[aps,prb,twocolumn,superscriptaddress]{revtex4-2}

\usepackage{color}
\usepackage{graphicx}
\usepackage{dcolumn}
\usepackage{bm}
\usepackage{float}
\usepackage{hyperref}
\usepackage[mathlines]{lineno}
\usepackage{tabularx}
\usepackage{hyperref}
\usepackage{footnote}
\usepackage{amsmath}
\usepackage{txfonts}

\begin{document}

\title{Stabilizing a hydrogen-rich superconductor at 1 GPa by the charge-transfer modulated virtual high-pressure effect}

\author{Miao Gao}\email{gaomiao@nbu.edu.cn}\affiliation{Department of Physics, School of Physical Science and Technology, Ningbo University, Zhejiang 315211, China}\affiliation{School of Physics, Zhejiang University, Hangzhou 310058, China}
\author{Peng-Jie Guo}\affiliation{Department of Physics, Renmin University of China, Beijing 100872, China}
\author{Huan-Cheng Yang}\affiliation{Department of Physics, Renmin University of China, Beijing 100872, China}
\author{Xun-Wang Yan}\affiliation{College of Physics and Engineering, Qufu Normal University, Shandong 273165, China}
\author{Fengjie Ma}\affiliation{The Center for Advanced Quantum Studies and Department of Physics, Beijing Normal University, Beijing 100875, China}
\author{Zhong-Yi Lu}\affiliation{Department of Physics, Renmin University of China, Beijing 100872, China}
\author{Tao Xiang}\affiliation{Institute of Physics, Chinese Academy of Sciences, Beijing 100190, China }
\affiliation{School of Physical Sciences, University of Chinese Academy of Sciences, Beijing 100049, China}
\affiliation{Beijing Academy of Quantum Information Sciences, Beijing 100193, China}
\author{Hai-Qing Lin}\email{hqlin@zju.edu.cn}\affiliation{School of Physics, Zhejiang University, Hangzhou 310058, China}

\date{\today}

\begin{abstract}

Applying pressure around megabar is indispensable in the synthesis of high-temperature superconducting hydrides, such as H$_3$S and LaH$_{10}$.
Stabilizing the high-pressure phase of hydride around ambient condition is a severe challenge.
Based on the density-functional theory calculations, we give the first example that the structure of hydride CaBH$_5$ predicted above 280 GPa, can maintain its dynamical stability with pressure down to 1 GPa,
by modulating the charge transfer from metal atoms to hydrogen atoms via the replacement of Ca with alkali metal atoms e.g. Cs, in which the
[BH$_5$]$^{2-}$ anion shrinks along $c$ axis and expands in the $ab$ plane, experiencing an anisotropic virtual high pressure.
This mechanism, namely charge transfer modulated virtual high pressure effect, plays a vital role in enhancing the structural stability and leading to the reemergence of ambient-pressure-forbidden [BH$_5$]$^{2-}$ anion around 1 GPa in CsBH$_5$.
Moreover, we find that CsBH$_5$ is a strongly coupled superconductor, with transition temperature as high as 98 K, well above the liquid-nitrogen temperature.
Our findings provide a novel mechanism to reduce the critical pressure required by hydrogen-rich compound without changing its crystal structure, and also shed light on searching ambient-pressure high-temperature superconductivity in metal borohydrides.

\end{abstract}

\maketitle
\section{Introduction}

Since the discoveries of high-temperature superconductivity in H$_3$S and LaH$_{10}$ at ultrahigh pressures \cite{Drozdov-Nature525,Drozdov-Nature569,Somayazulu-PRL122}, lots of efforts have been devoted to reduce the critical pressure required by hydrogen-rich compounds.
However, a limited number of hydrides were claimed to be high-temperature superconductors at relatively lower pressures after the exploration of diagram of binary hydrides \cite{Duan-NSR4,Oganov-NRM4,Zurek-JCP150,Livas-PR856,Hutcheon-PRB101,Shipley-PRB104,Chen-PRL127}.
For instance, RbH$_{12}$ \cite{Hutcheon-PRB101} and NaH$_6$ \cite{Shipley-PRB104} were predicted to be stable around 50 GPa and 100 GPa with transition temperature ($T_c$) being 115 K and 279 K, respectively, based on density-functional theory (DFT) calculations.
It was reported that CeH$_{10}$ and CeH$_9$ show superconducting transitions around 115 K and 57 K below 100 GPa \cite{Chen-PRL127}.
Owing to the enlarged phase space and enhanced flexibility, it is promising to find high-$T_c$ superconductors under low pressure condition in ternary hydrides.

The studies of superconductivity in ternary hydrides can be classified into two categories. (I) Chemically doping the binary superconducting hydrides that are either synthesized or predicted.
The influence of partial substitution of sulphur atom on superconductivity of H$_3$S was carried out through DFT computations \cite{Ge-PRB93}.
The thermodynamical stability and superconductivity of H$_6$S$X$ ($X$=Cl and Br) were investigated \cite{Hai-PRB105}. As suggested, H$_6$SCl is a candidate of high-$T_c$ superconductor with $T_c$ about 155 K at 90 GPa.
Several metastable phases of CSH$_7$ stoichiometry, H$_3$S with CH$_4$ intercalation, are potential superconductors, with $T_c$ ranging from 100 K to 190 K above 100 GPa \cite{Cui-PRB101}.
By introducing additional electrons into MgH$_{16}$ \cite{Lonie-PRB87}, Li$_2$MgH$_{16}$ was anticipated to show superconductivity above 470 K at 250 GPa \cite{Sun-PRL123}.
LaBeH$_8$ and LaBH$_8$ were predicted to exist below 50 GPa, meanwhile keeping high-$T_c$ states \cite{Zhang-PRL128,Cataldo-PRB104,Liang-PRB104}, since additional chemical pressure is generated through insertion of small radius atoms into the cubic H$_8$ units of UH$_8$-like structure.
A series of ternary alloys hydrides, including La-Y-H \cite{Semenok-MT48}, La-Ce-H \cite{Chen-arXiv,Bi-arXiv}, and La-Nd-H \cite{Semenok-arXiv} compounds, have been synthesized with superconducting properties similar to their binary analogues. The Y-Ca-H phase diagram was searched theoretically \cite{Zhao-PRB106}.

(II) By metallizing the H-related $\sigma$-bonding bands. This establishes a strong link between superconducting hydrides and the famous phonon-mediated high-$T_c$
superconductor MgB$_2$, and offers a clear clue to guide the exploration of new type high-$T_c$ hydrides.
In fact, H$_3$S is a MgB$_2$-type superconductor with a higher characteristic frequency, acting as the energy scale in pairing electrons \cite{Bernstein-PRB91}.
To fulfill this strategy, we have elucidated that boron is an ideal bonding partner of hydrogen \cite{Gao-PRB104}. The physical reasons lie in two aspects, i.e., large bonding strength between boron and hydrogen, and slightly smaller electronegativity of boron with respect to that of hydrogen.
In particular, the validity of this proposal was unambiguously proved by using KB$_2$H$_8$ as an outstanding example, formed by inserting anion [BH$_4$]$^{-}$ into the face-centered-cubic lattice of potassium. We found that KB$_2$H$_8$ can become a superconductor with $T_c$ about 140 K at 12 GPa. Certainly, similar properties will be acquired by replacing K with other alkali metal elements, for instance Rb and Cs,
as reported in a subsequent calculation \cite{Li-PRB105}. Motivated by our KB$_2$H$_8$ work, the electron-phonon coupling (EPC) and superconductivity in isostructural $M$C$_2$H$_8$ ($M$=Na, K, Mg, Al, Ga) \cite{Jiang-PRB105} and $M$N$_2$H$_8$ ($M$=Be, Mg, Al, Cu, Zn, etc.) \cite{Wan-arXiv} were investigated. Recently, it was shown that doping holes into $\alpha$-Ca(BH$_4$)$_2$, a [BH$_4$]$^{-}$ containing molecular crystal, may result in room-pressure superconductivity above 110 K, originated from the strong EPC between
B-H $\sigma$ molecular orbitals and bond-stretching phonons \cite{Cataldo-arXiv}.

These studies have demonstrated that metal borohydrides supply a promising playground to hunt high-$T_c$ superconductors near ambient pressure.
Further exploration of novel B-H anions and their potential in constructing superconducting borohydrides are desired.
Although, there are several known B-H anions \cite{Deiseroth-ZAAC571,Kuznetsov-RJIC32,Tiritiris-ZAAC629,Ohba-PRB74,Zhang-PRB82}, such as [B$_3$H$_8$]$^-$, $closo$-type dianions [B$_n$H$_n$]$^{2-}$ ($n$=5-12), and [B$_2$H$_6$]$^{2-}$. The existence of B-B bonds inside these anions brings about the domination of B-B bonding states around the valance band maximum \cite{Ohba-PRB74}.
This unlikely induces high-$T_c$ superconductivity, since hydrogen-related electronic states can not participate effectively in EPC via doping holes.
Therefore, the absence of B-B bonds in B-H anion should be regarded as a principal selection rule.
Particularly, trivalent [BH$_6$]$^{3-}$ and divalent [BH$_5$]$^{2-}$ uncovered by delving the ternary Li-B-H and Ca-B-H phase diagrams under pressures \cite{Kokail-PRM1,Cataldo-PRB102}, prefectly satisfy the selection rule. Li$_2$BH$_6$ was suggested to superconduct around 100 K under 100 GPa. In sharp contrast, the insulating hexagonal CaBH$_5$ [Fig.~\ref{fig:Stru}] is stable above 280 GPa. An insulator-metal transition occurs at about 300 GPa, but no superconductivity. Here, we focus on the [BH$_5$]$^{2-}$ anion, since it is feasible to turn insulating CaBH$_5$ into metallic, and a potential MgB$_2$-like high-$T_c$ superconductor, by replacing Ca with monovalent metal atoms.

In this work, the dynamical stability of $M$BH$_5$ ($M$=Ca, Li, Na, K, Rb, Cs) and superconductivity near ambient pressure are systematically investigated based on DFT calculations (for details of the calculations, see the supplemental material \cite{Supp} and 15 references therein).
At room pressure, CaBH$_5$ remains insulating, but there are substantial imaginary modes in the phonon spectrum. Generally, the decline in occupation number of valance band of an insulator,
weakens the chemical bonding and easily triggers a structural phase transition. Intuitively, substituting alkali metal for calcium significantly improves the dynamical stability.
The geometry of [BH$_5$]$^{2-}$ anion shows anisotropic variants along the $c$ axis and in the $ab$ plane, governed by charge transfer from metal atoms.
In particular,
all imaginary phonon modes disappear in CsBH$_5$ around 1 GPa.
This provides a completely new avenue to stabilize high-pressure hydrides around ambient condition, without destroying the crystal structure.
Different from metallic CaBH$_5$, whereas the electronic density of states (DOS) at the Fermi level, $N(0)$, comes from Ca-$3d$ orbitals,
$N(0)$ is mainly contributed by H orbitals for CsBH$_5$, a key ingredient for high-$T_c$ hydrides.
Utilizing Wannier interpolation technique, the EPC and properties of superconductivity in CsBH$_5$ are accurately computed.
We find that the EPC is quite strong, with $\lambda$ being 3.96 at 1 GPa, about 1.8 times those in H$_3$S and LaH$_{10}$ \cite{Duan-SR4,Liu-PNAS114}.
By solving the anisotropic Eliashberg equations, the $T_c$ of CsBH$_5$ is predicted to be 83-98 K from 1 GPa to 7GPa.
This makes CsBH$_5$ a seldom reported hydrogen-rich superconductor above the liquid-nitrogen temperature around ambient pressure.

\section{Results and discussions}

\begin{figure}[tbh]
\begin{center}
\includegraphics[width=8.6cm]{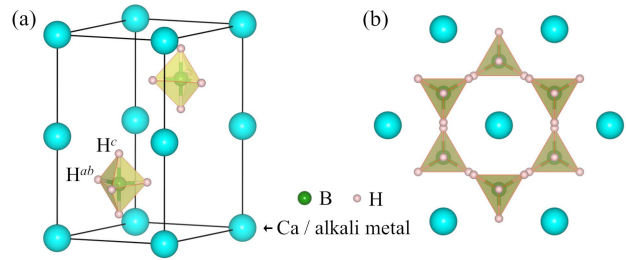}
\caption{(a) Three-dimensional view of $M$BH$_5$ ($M$=Ca, or alkali metal atoms). (b) Top view.
Here, the lattice parameters of CsBH$_5$ at ambient pressure are adopted to draw the crystal structure.
The cyan, green, and pink balls represent calcium/alkali metal, boron, and hydrogen atoms, respectively. The solid black line denotes the unit cell. Two kinds of nonequivalent hydrogen atoms are denoted by H$^{c}$ and H$^{ab}$.}
\label{fig:Stru}
\end{center}
\end{figure}

The crystal structure of $M$BH$_5$ is
shown in Fig. \ref{fig:Stru}, whereas [BH$_5$]$^{2-}$ anions are sandwiched between two neighboring triangular layers of $M$ atoms.
Boron atom locates at the center of [BH$_5$]$^{2-}$, with five satellite-like hydrogen atoms. Hydrogen atoms along the $c$ axis and in the $ab$ plane are nonequivalent, which are labelled as H$^{c}$ and H$^{ab}$.
$M$BH$_5$ belongs to the space group $P6_3$/$mmc$ (No. 194), with $M$, boron, H$^{c}$, and H$^{ab}$ atoms occupying the $2a$ (0.0, 0.0, 0.0), $2d$ ($1/3$, $2/3$, $1/4$), $4f$ ($1/3$, $2/3$, $x$)
and $6h$ ($y$, 1-$y$, $1/4$) Wyckoff positions, respectively.
After optimization, we find that $x$=0.4492, and $y$=0.5089 for CaBH$_5$ at ambient pressure. The lattice constants are computed as $a$=3.9669 {\AA}, and $c$=8.4350 {\AA}.

\begin{figure}[tbh]
\begin{center}
\includegraphics[width=8.6cm]{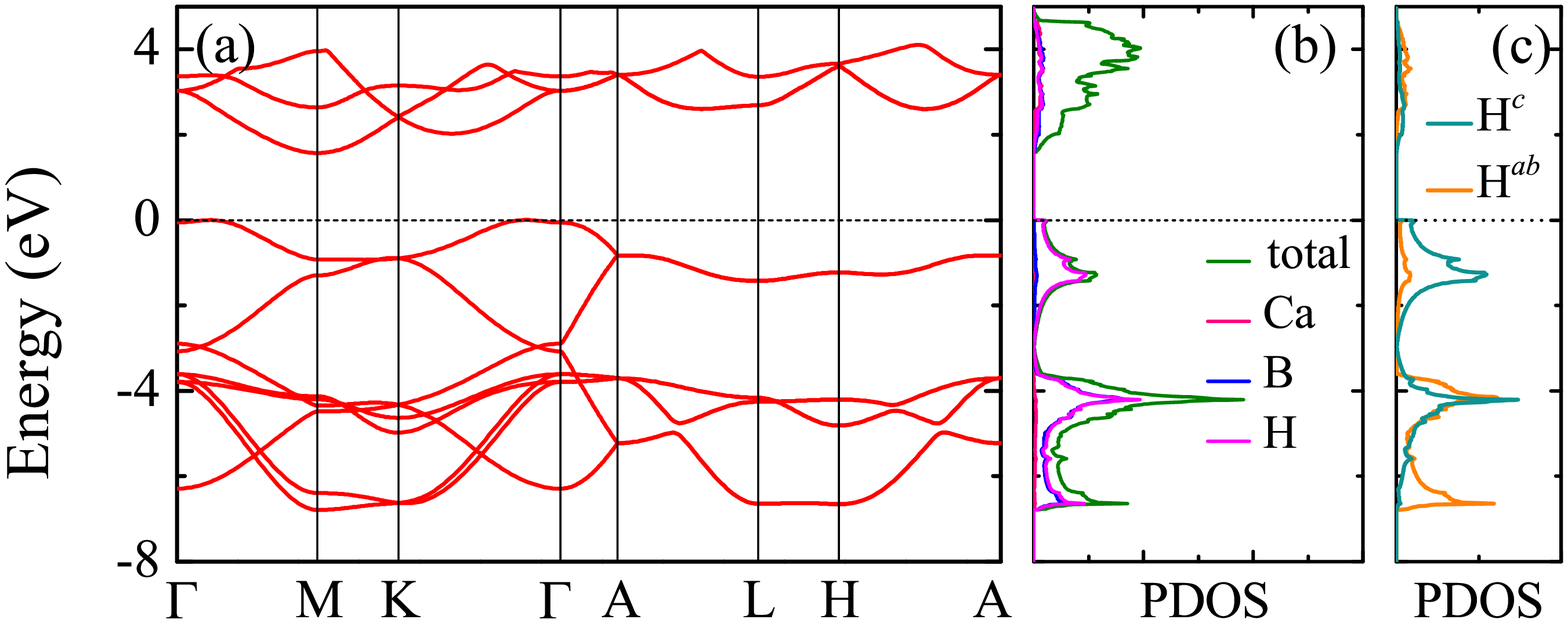}
\includegraphics[width=8.6cm]{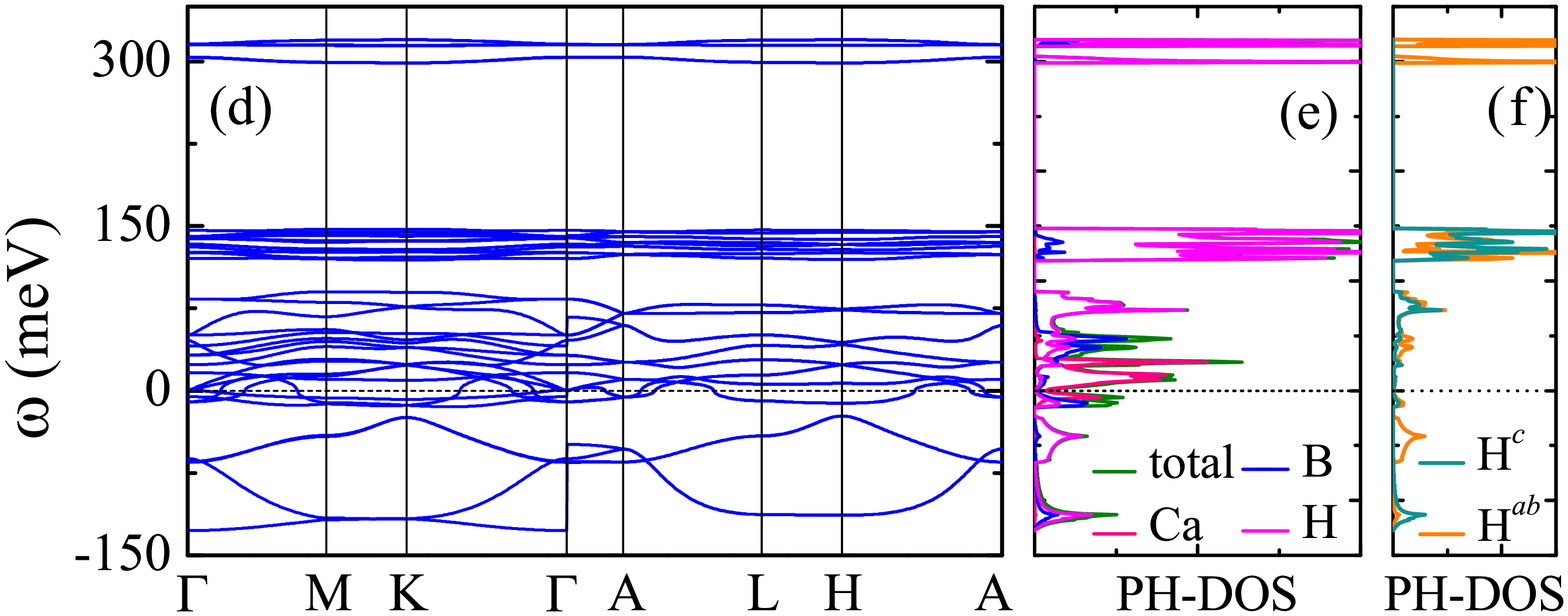}
\caption{Electronic structure and lattice dynamics of CaBH$_5$ at ambient pressure. (a) Band structure. (b) Total and partial DOS. (c) DOS projected onto two nonequivalent hydrogen atoms. (d) Phonon spectrum. (e) Total and projected phonon DOS. (f) Phonon DOS from two types of hydrogen atoms.}
\label{fig:Ca-Band}
\end{center}
\end{figure}

Figure \ref{fig:Ca-Band} presents the electronic structure and lattice dynamics of room-pressure CaBH$_5$. CaBH$_5$ is an insulator, with an indirect band gap of 1.58 eV [Fig. \ref{fig:Ca-Band}(a)].
As shown by the partial DOS [Fig. \ref{fig:Ca-Band}(b)], there are marked overlaps between B and H orbitals from -7.0 eV to -3.0 eV, suggesting strong chemical bonding between them.
The electronic states of valence band mostly come from the H$^{c}$ orbitals [Fig. \ref{fig:Ca-Band}(c)]. However, CaBH$_5$ is unstable at ambient condition [Fig. \ref{fig:Ca-Band}(d)], as revealed by substantial imaginary phonon modes \cite{Im}.
Here, we introduce a quantity $\bar{\omega}^i=\int_{-\infty}^{0}\omega F(\omega) d\omega/\int_{-\infty}^{0}F(\omega) d\omega$, corresponding to the average value of imaginary phonon frequencies in the entire Brillouin zone, to measure the dynamical stability. $F(\omega)$ is the phonon DOS. For CaBH$_5$, $\bar{\omega}^i$=-53.23 meV.
The imaginary modes mainly involve the vibration of H atoms [Fig. \ref{fig:Ca-Band}(e) and Fig. \ref{fig:Ca-Band}(f)].
Compared with the cell parameters under 300 GPa, the physical reason for instability of CaBH$_5$ when quenched to ambient pressure becomes clear.
At 300 GPa, the lattice constants are $a$=3.0947 {\AA}, and $c$=4.8110 {\AA}.
Owing to short $c$ axis, the Coulomb repulsion between Ca atoms along the longitudinal direction is quite strong. As a consequence, the $c$ axis will expand dramatically after the removal of pressure.
On the other hand, boron can only supply three valence electrons, H$^{c}$ atoms trend to relax towards Ca atoms to acquire more electrons.
At last, the H$^c$ atom almost locates in the same plane of Ca layer, with the Wyckoff parameter $x$ being 0.5040.
The combined effects result in an unphysical B-H$^{c}$ bond length, 1.6804 {\AA},
in comparison with the typical value ($\sim$1.22 {\AA}) of B-H bonds at ambient pressure \cite{Cataldo-PRB102}.
Hence, the emergence of imaginary modes can be rationalized in terms of the abnormal B-H$^c$ bond length.

\begin{figure}[t]
\begin{center}
\includegraphics[width=8.6cm]{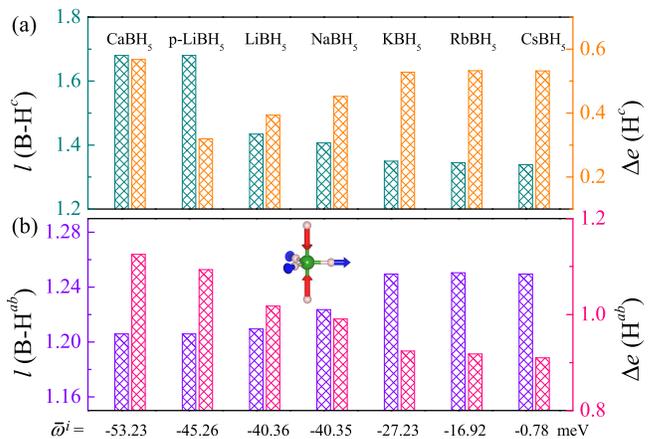}
\caption{(a) Length of B-H$^c$ bond and charge obtained by H$^c$ atom.
(b) Cases for H$^{ab}$ atom.
A cartoon of [BH$_5$]$^{2-}$ behavior is given as insert in (b). The average values of imaginary phonon freqencies $\bar{\omega}^i$ for $M$BH$_5$ ($M$=Ca, Li, Na, K, Rb, and Cs) at ambient pressure, are listed at the bottom. $p$-LiBH$_5$ is the $pesudo$-LiBH$_5$ crystal described in the main text. CsBH$_5$ is stabilized at 1 GPa, as shown in Fig. \ref{fig:phonon}. }
\label{fig:charge}
\end{center}
\end{figure}

The changing rule of [BH$_5$]$^{2-}$ anion, after the substitution of alkali metal for Ca atoms, is summarized in Fig. \ref{fig:charge}.
Because alkali metal has just one valence electron, H$^c$ can not obtain sufficient electrons by approaching the metal atoms.
To verify this, we construct a suppositional crystal, in which the lattice constants and inner coordinates of room-pressure CaBH$_5$ are adopted, while Ca is replaced with Li, namely $pseudo$-LiBH$_5$. As expected, H$^c$ acquires only 0.24 electrons in $pseudo$-LiBH$_5$, evidently smaller than 0.57 in CaBH$_5$, through the Bader charge analysis \cite{Henkelman-CMS36,Tang-JPCM21}.
H$^c$ atoms have no choice but to fight over H$^{ab}$ for electrons from boron, causing shortened B-H$^c$ bonds.
Inevitably, the charge transfer from B to H$^{ab}$ is depressed. Meanwhile, B-H$^{ab}$ bonds are elongated after withstanding the competition from H$^c$ atoms [Fig. \ref{fig:charge}(b)].
This trend is unambiguous by comparing the data of $pseudo$-LiBH$_5$ with those in optimized LiBH$_5$.
From Li to Cs, the unit cell gradually dilates, as well as the $M$-H$^c$ distance.
This further prevents H$^c$ from obtaining electrons of $M$, and accelerates the process of H$^c$ relaxing towards B.
As indicated by $\bar{\omega}^i$, the imaginary modes are gradually ameliorated, and finally disappear in CsBH$_5$ at 1 GPa.
Under 1 GPa, the bond lengths of B-H$^c$ and B-H$^{ab}$ in CsBH$_5$ are 1.3290 {\AA} and 1.2507 {\AA}.
If CaBH$_5$ has the same bond lenghts, the lattice constants $a$ and $c$ must be artificially adjusted to 4.11 {\AA} and 5.02 {\AA}, resulting in
13.7 GPa and 70.4 GPa pressures in the $ab$ plane and along the $c$ axis, respectively.
From this viewpoint, [BH$_5$]$^{2-}$ anion in CsBH$_5$ undergoes anisotropic virtual high-pressure effect, caused by lessening charge transfer from metal atom.
Different from the concept of chemical precompression raised by N. W. Ashcroft \cite{Ashcroft-PRL92}, we call it charge transfer modulated virtual high pressure effect, which offers a novel mechanism
to stabilize the high-pressure hydrogen-rich compound near ambient condition.
To the best of our knowledge, this mechanism has never been proposed.

Moreover, we find that CsBH$_5$ is dynamically stable with pressure up to 7 GPa.
We carry out the structure search using the Universal Structure Predictor: Evolutionary Xtallography method (USPEX) at an intermediate pressure, i.e., 4 GPa. It suggests that the target crystal studied here is a metastable state \cite{Supp}.
The static formation enthalpies of CsBH$_5$ are found to be -0.340, -0.811, -1.042, and -1.123 eV/f.u. under pressures of 1, 3, 5, and 7 GPa, with respect to $\alpha$-B$_{12}$, $P6_3/m$ hydrogen \cite{Pickard-NP3}, and the ground state of caesium. Here, the body-centered cubic, face-centered cubic, and tetragonal $I4_1/amd$ structures of caesium are employed, according to its phase transition sequence under pressure \cite{Takemura-PRL49}. We include the zero-point correction to static formation enthalpy using phonon frequencies at the $\Gamma$ point. As a consequence, the formation enthalpies become 0.178, -0.560, -0.741, and -0.898 eV/f.u., respectively. We also compare the formation enthalpy of CsBH$_5$ versus the combination of $\frac{1}{2}$B$_2$H$_6$+CsH+$\frac{1}{2}$H$_2$ \cite{atomly}. The relative formation enthalpies are equal to  0.400, 0.137, -0.001, and -0.053 eV/f.u. at corresponding pressures.
These results indicate that the synthesis of CsBH$_5$ is not impossible.
Since CsBH$_5$ is stable around ambient pressure, it is interesting to calculate the electronic structure, lattice dynamics, and superconductivity.

\begin{figure}[t]
\begin{center}
\includegraphics[width=8.6cm]{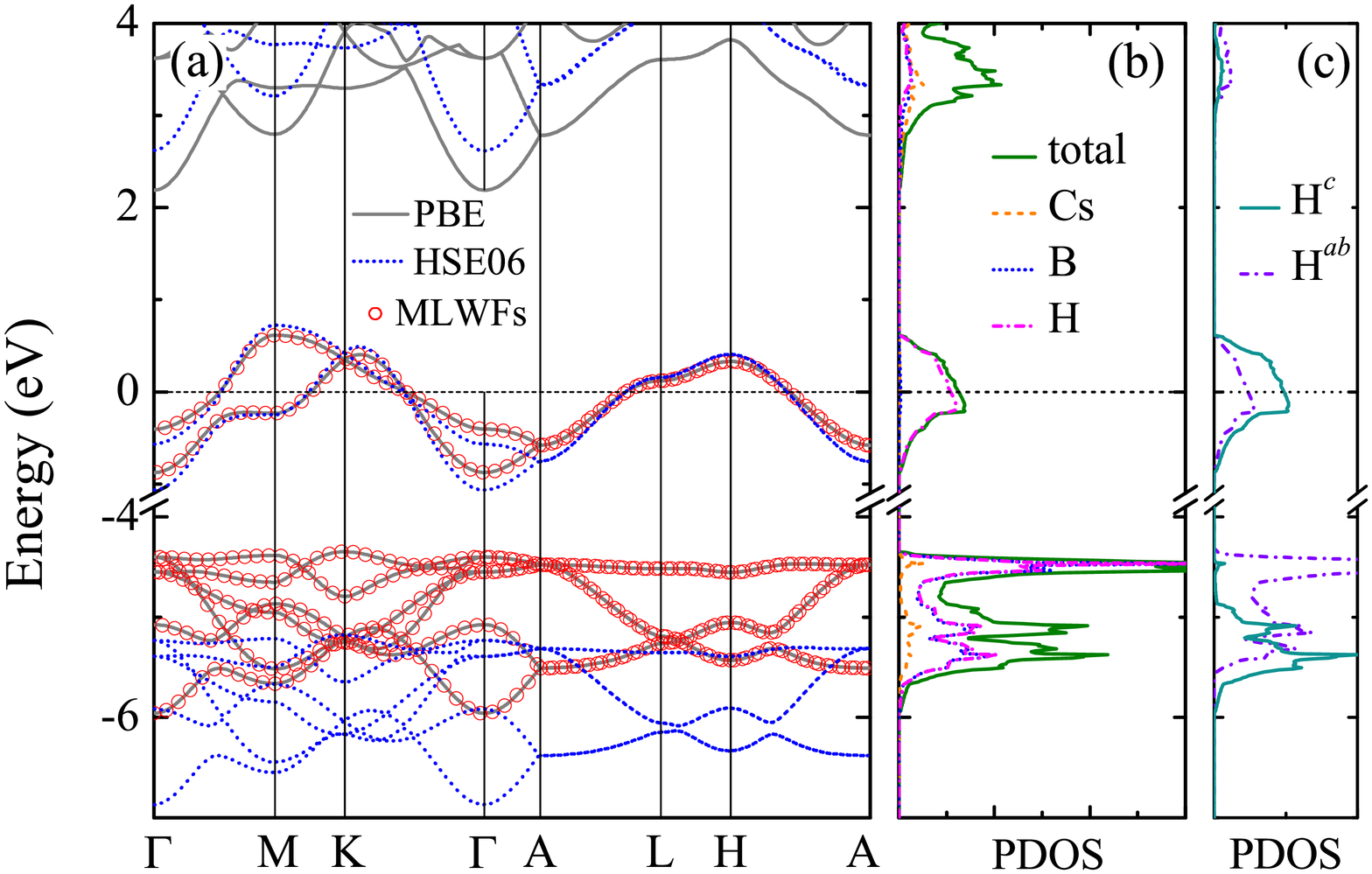}
\includegraphics[width=8.6cm]{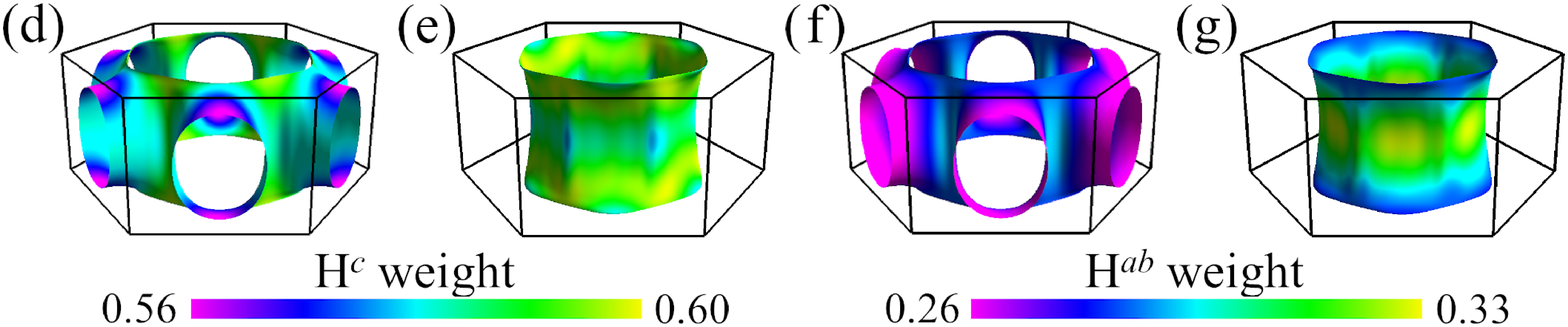}
\includegraphics[width=8.6cm]{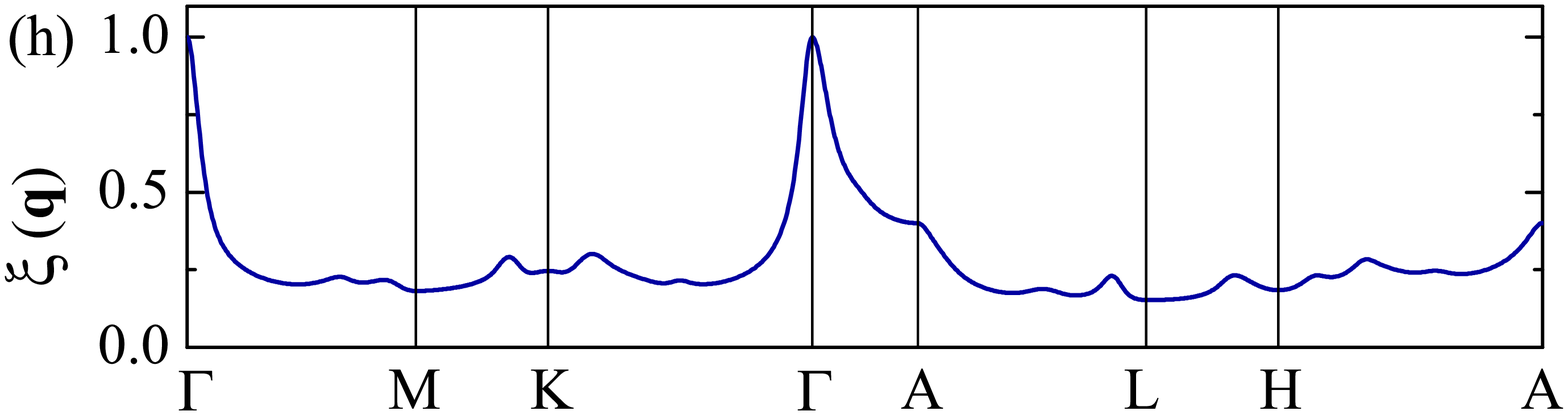}
\caption{Electronic structure of CsBH$_5$ under 1 GPa. (a) Band structure. The gray lines and red circles stand for the bands obtained by the first-principles calculation and the maximally localized Wannier functions (MLWFs) interpolation, respectively. The results generated by HSE06 functional are shown as dotted blue lines. The Fermi level is set to zero. (b) Total and partial DOS. (c) DOS contributed by two kinds of hydrogen atoms.
Spectral weight on the Fermi surfaces for (d-e) the H$^c$ and (f-g) the H$^{ab}$ atoms, respectively.
(h) Fermi surface nesting $\xi({\bf q})$, normalized by $\xi({\Gamma})$}.
\label{fig:Band}
\end{center}
\end{figure}

Figure \ref{fig:Band} shows the electronic structure of CsBH$_5$ at 1 GPa.
There are two energy bands across the Fermi level [Fig. \ref{fig:Band}(a)].
The band structures generated through MLWFs interpolation are perfectly in accordance with the DFT results.
This establishes a solid foundation for subsequent Wannier interpolation in the computation of EPC.
These two partially filled bands are degenerate along the high-symmetry $A$-$L$-$H$-$A$ line.
As revealed in Fig. \ref{fig:Band}(b) and Fig. \ref{fig:Band}(c), DOS around the Fermi level is dominated by H orbitals, especially H$^c$.
Specifically, H$^c$ and H$^{ab}$ occupy 58.4\% and 28.3\% of $N(0)$.
Compared with the case of CaBH$_5$ [Fig. \ref{fig:Ca-Band}(c)], the energies of H$^c$ and H$^{ab}$ orbitals shift downwards and upwards, respectively,
consistent with the charge transfer picture.
Since PBE always underestimates the band gap, we recalculate the band structure with HSE06 hybrid functional. The filled bands almost rigidly shift downwards by 0.8 eV.
The influence of hybrid functional on the metallic bands is insignificant.
The electronic states possessing large proportion of H$^{c}$ orbitals mainly locate near the planes with $k_z=\pm\pi$ [Fig. \ref{fig:Ca-Band}(d) and Fig. \ref{fig:Ca-Band}(e)].
On the contrary, H$^{ab}$ orbitals contribute one third to states on the cylindrical Fermi surface with $k_z$ near zero [Fig. \ref{fig:Ca-Band}(f) and Fig. \ref{fig:Ca-Band}(g)].
The peak of Fermi surface nesting function is along $\Gamma$-$A$ direction, owing to the cylindrical Fermi surface [Fig. \ref{fig:Ca-Band}(h)].

\begin{figure}[t]
\begin{center}
\includegraphics[width=8.6cm]{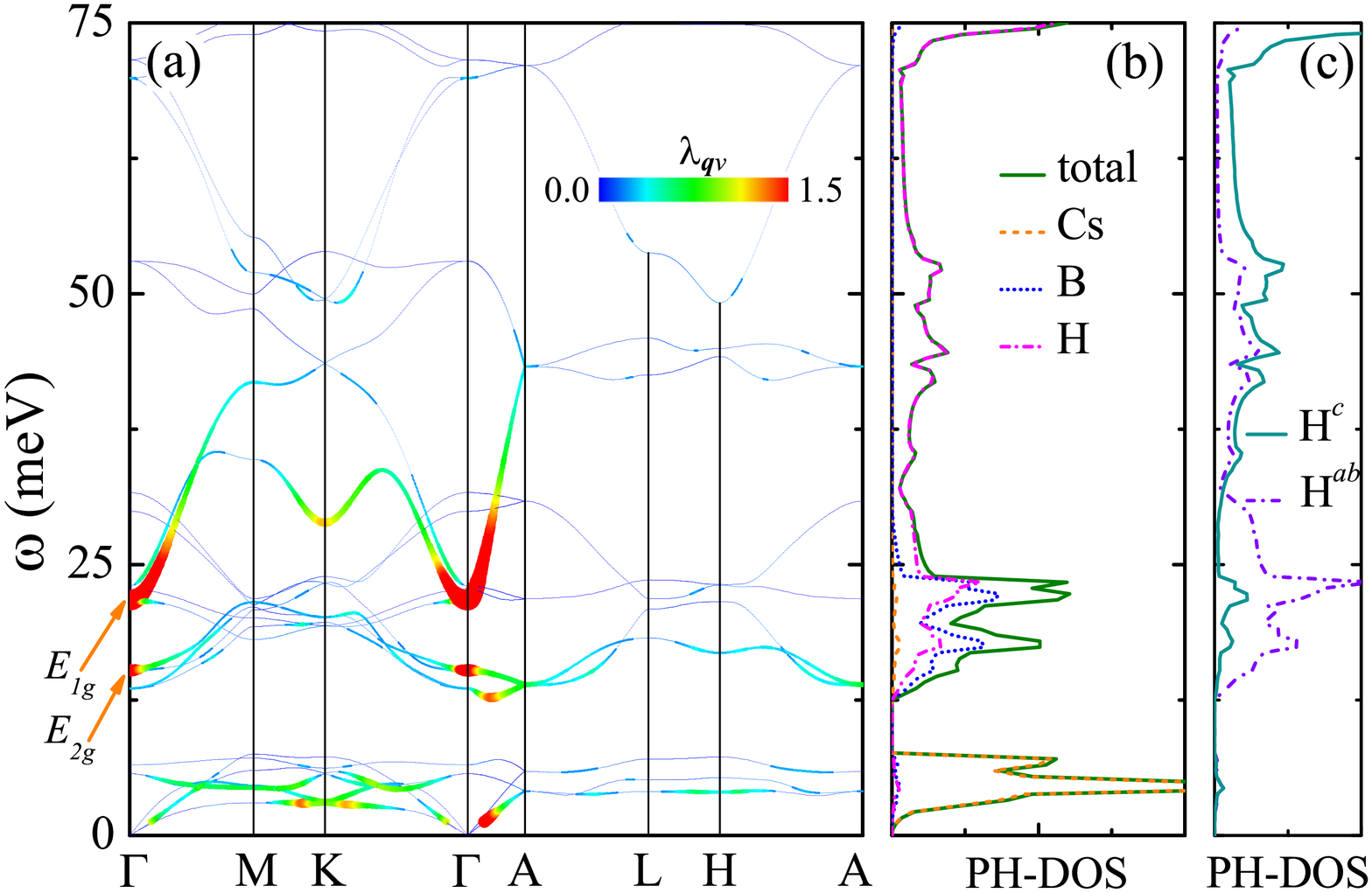}
\caption{Lattice dynamics of CsBH$_5$ at 1 GPa. (a) Phonon spectrum below 100 meV.
Both the mapped color and the line width represent the strength of wavevector-resolved EPC constant $\lambda_{{\bf q}v}$. For the entire spectrum, please see Fig. S9 \cite{Supp}. (b-c) Total and projected phonon DOS. }
\label{fig:phonon}
\end{center}
\end{figure}

The phonon spectrum of CsBH$_5$ at 1 GPa is shown in Fig. \ref{fig:phonon}(a).
The lattice stability is verified by the fact that no imaginary phonon modes appear in the spectral function.
Below 30 meV, there are several strongly coupled phonon modes, the mode symmetries are identified as $E_{1g}$ and $E_{2g}$ at the $\Gamma$ point.
From 10 meV to 30 meV, phonon modes mix the movements of boron and hydrogen atoms together, according to projected phonon DOS [Fig. \ref{fig:phonon}(b)].
However, the spectral weight of H$^c$ is negligible below 30 meV [Fig. \ref{fig:phonon}(c)]. Cs atoms mainly participate in the phonon modes with frequencies lower than 10 meV, which also have sizeable EPC. Thus, phonons associated with Cs, B, and H$^{ab}$ atoms possess strong coupling with electrons, since their movements affect the charge transfer, as well as the electronic states around the Fermi level. Interestingly, the vibration of H$^c$ atoms has insignificant contribution to EPC, although, they dominate the DOS around the Fermi level.
The band structures and phonon spectra of $M$BH$_5$ ($M$=Li, Na, K, Rb, and Cs) under ambient pressure are given in Fig. S4-Fig. S8 \cite{Supp}.

\begin{figure}[tbh]
\begin{center}
\includegraphics[width=8.6cm]{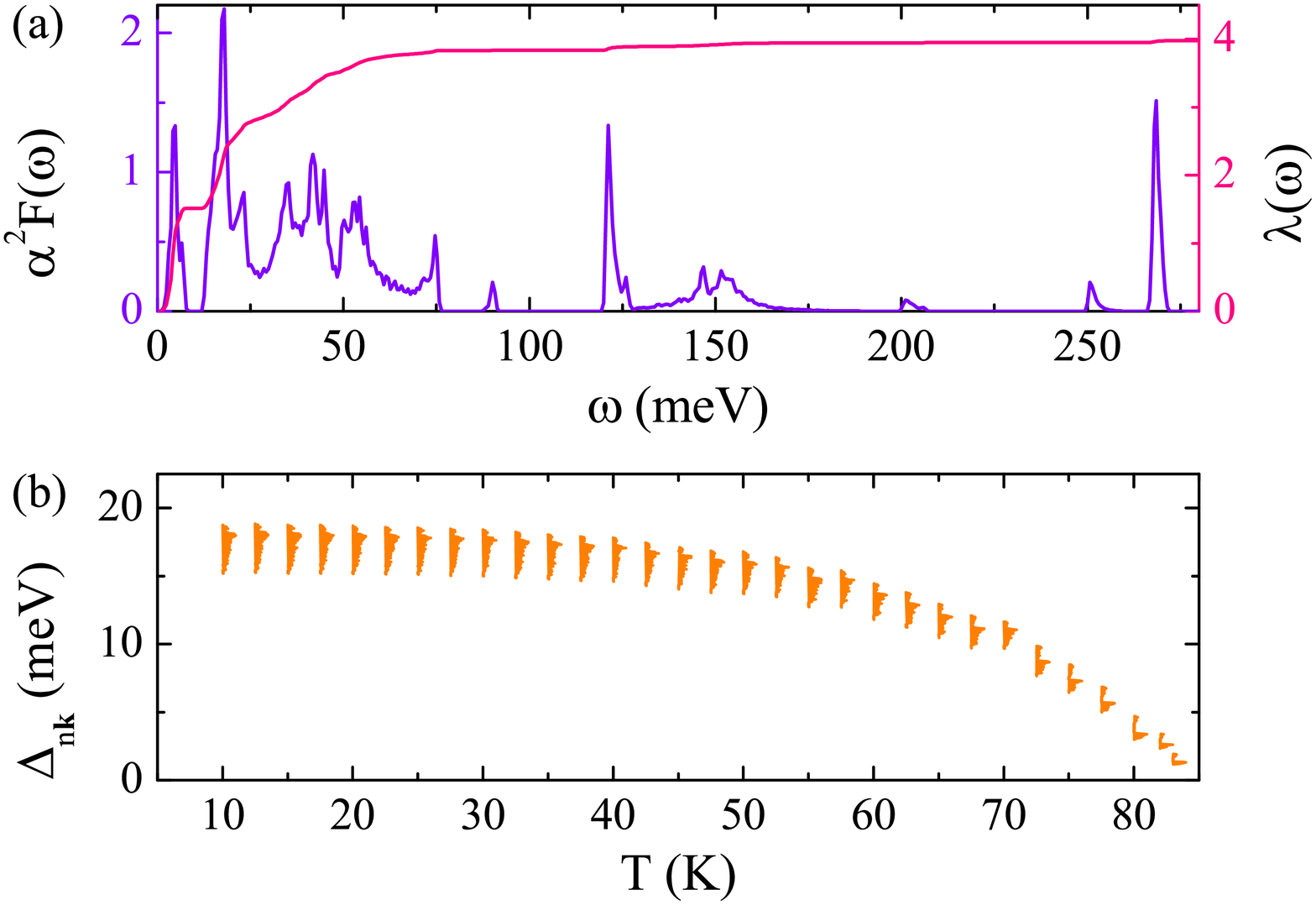}
\includegraphics[width=8.6cm]{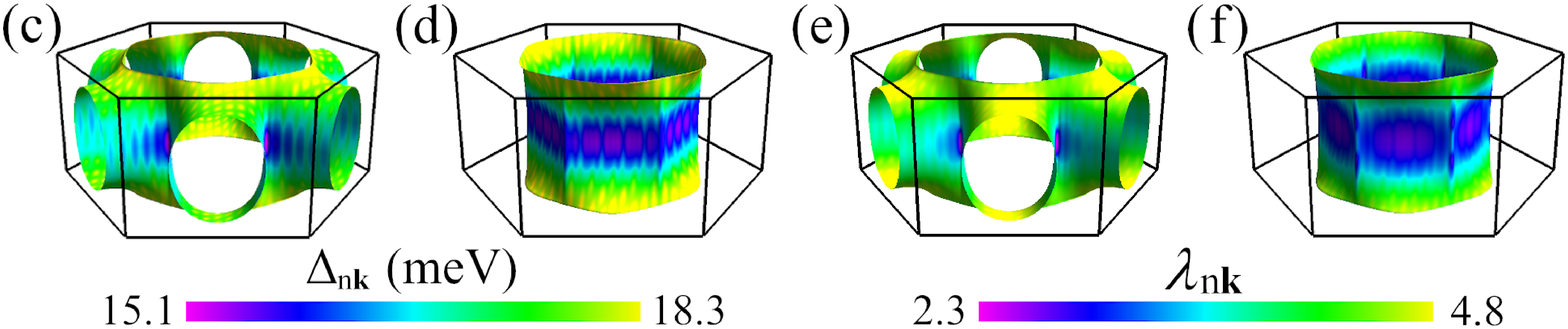}
\caption{(a) Eliashberg spectral function $\alpha^2F(\omega)$ and the accumulated EPC $\lambda(\omega)$ for CsBH$_5$ under 1 GPa.
(b) The superconducting energy gaps $\Delta_{n\bf k}$ at different temperatures for $\mu^*$=0.10. (c-d) The distribution of $\Delta_{n\bf k}$ on the Fermi surfaces at 10 K.
(e-f) The distribution of momentum-resolved EPC constant $\lambda_{n\bf k}$ on the Fermi surfaces. }
\label{fig:a2f}
\end{center}
\end{figure}

Figure \ref{fig:a2f} shows the Eliashberg spectral function $\alpha^2F(\omega)$ and the superconducting energy gaps $\Delta_{n\bf{k}}$ at different temperatures for CsBH$_5$ at 1 GPa.
By integrating $\alpha^2F(\omega)$, the EPC constant $\lambda$ and the logarithmic average frequency $\omega _{\text{log}}$ are equal to 3.96 and 13.95 meV, respectively.
The coupling strength in CsBH$_5$ is markedly stronger than 2.19 in H$_3$S at 200 GPa \cite{Duan-SR4} and 2.20 in LaH$_{10}$ at 250 GPa \cite{Liu-PNAS114}.
According to the accumulated EPC $\lambda(\omega) = 2\int_0^\omega \frac{1}{\omega'} \alpha^2F(\omega') d\omega'$, phonons below 30 meV contribute 73.2\% to $\lambda$, especially 38.2\% is originated from Cs phonons [Fig. \ref{fig:a2f}(a)]. Consequently, $\omega _{\text{log}}$, the energy scale of pairing interaction between electrons, is relatively small in CsBH$_5$.
If we set the Coulomb pseudopotential $\mu^*$ to its common value 0.10, the highest temperature with non-vanished superconducting gaps, namely $T_c$, is found to be 83 K [Fig. \ref{fig:a2f}(b)], successfully exceeding the liquid-nitrogen temperature, by solving the anisotropic Eliashberg equations.
At each temperature, the superconducting gaps group together, resulting in a single-gap characteristics.
At 10 K, the superconducting energy gaps $\Delta_{n\bf{k}}$ exhibit visible anisotropy [Fig. \ref{fig:a2f}(c) and Fig. \ref{fig:a2f}(d)], with the mean value being 17.27 meV. Therefore, CsBH$_5$ is an anisotropic single-gap $s$-wave superconductor. The anisotropic behavior of $\Delta_{n\bf{k}}$ is mainly determined by the distribution of $\lambda_{n\bf{k}}$ on the Fermi surfaces [Fig. \ref{fig:a2f}(e) and Fig. \ref{fig:a2f}(f)], which resembles that of H$^c$ orbitals [Fig. \ref{fig:Band}(d) and Fig. \ref{fig:Band}(e)]. Thus, we can conclude that H$^c$ orbitals have strong coupling with vibrations of Cs, B, and H$^{ab}$ atoms.

The superconductivity of CsBH$_5$ under higher pressures is further investigated [Fig. S10].
We find that there is a monotonic rise in $T_c$ with the increase of pressure.
Interestingly, the dependence of $T_c$ on pressure is perfectly linear, and the slope $dT_c/dP$ is calculated to be 2.5 K/GPa, leading to the highest value of 98 K at 7 GPa.
If we set the Coulomb pseudopotential $\mu^*$ to its upper bound, commonly taken as 0.15, the $T_c$ of CsBH$_5$ is 78-89 K from 1 GPa to 7 GPa, still above the liquid-nitrogen temperature.
The bond lengths are shortened, resulting in the aggrandizement of atomic force constants and hardening of phonons, e.g. $E_{1g}$ and $E_{2g}$ modes. For example, the frequencies of $E_{1g}$ and $E_{2g}$ increase to 46.52 meV and 27.22 meV at 7 GPa, respectively. However, the logarithmic average frequency $\omega _{\text{log}}$ does not simply follow this rule, especially $\omega _{\text{log}}$ is calculated to be 13.75 meV at 7 GPa, almost equal to that of 1 GPa. This means that, besides $E_{1g}$ and $E_{2g}$, new-type strongly coupled phonon modes emerge, particularly, the acoustic modes along $H$-$A$ line. These modes exhibit palpable softening upon the increase of pressure and play important roles in the enhancement of $T_c$.

\section{Conclusion}

In summary, we have proposed a charge transfer mechanism to stabilize the high-pressure phase of hydrogen-rich compound around ambient condition.
A virtual high-pressure effect resulting from the charge transfer is clearly elucidated in alkali metal substituted CaBH$_5$ through
DFT calculations. Physically, heavy hole doping in insulators is harmful to the stability.
Amazingly, introducing holes into CaBH$_5$ not only enhances the stability, but also induces high-$T_c$ phonon-mediated superconductivity.
This enriches the understanding of dynamical behavior of insulators under hole doping.
Furthermore, we find that dynamical stability can be achieved in CsBH$_5$ from 1 GPa to 7 GPa, with $T_c$ as high as 98 K.
It is noted that the existences of inequivalent hydrogen atoms in the borohydrides is of crucial importance for the success of charge transfer mechanism, as in the anisotropic [BH$_5$]$^{2-}$ anion studied here. This can be regarded as a criterion to select potential high-pressure metal borohydrides, whereas the charge transfer mechanism works.

\begin{acknowledgments}
This work was supported by the National Natural Science Foundation of China (Grant Nos. 11974194, 12088101, 11974207, 12074040, 11934020, and 11888101) and the National Key Research and Development Program of China (Grant No. 2017YFA0302900).
M.G. was also sponsored by K. C. Wong Magna Fund in Ningbo University.
\end{acknowledgments}

\end{document}